\documentclass[aps,twocolumn,prc,floatfix,showpacs,preprintnumbers,amsmath,amssymb,nofootinbib,groupedaddress]{revtex4-1}
\usepackage[normalem]{ulem}
\usepackage{wasysym}
\usepackage{color}
\usepackage{graphicx}
\usepackage{dcolumn}    
\usepackage{bigstrut}
\usepackage{amsmath,amsfonts,amsthm,bm}
\usepackage[pdfstartview=FitH,
            CJKbookmarks=true,
            bookmarksnumbered=true,
            bookmarksopen=true,
            colorlinks,
            pdfborder=001,
            linkcolor=blue,
            anchorcolor=blue,
            citecolor=blue
            ]{hyperref}

\begin{document}
%
\title{User guide for the {\tt hfbcs-qrpa} (v1) code}

\author{G.  Col\`o}
\email{colo@mi.infn.it}
\affiliation{Dipartimento di Fisica, Universit\`a degli Studi di Milano and INFN,
Sezione di Milano, Via Celoria 16, 20133 Milano, Italy.}

\author{X. Roca-Maza}
\email{xavier.roca.maza@mi.infn.it}
\affiliation{Dipartimento di Fisica, Universit\`a degli Studi di Milano and INFN,
Sezione di Milano, Via Celoria 16, 20133 Milano, Italy.}

\date{\today}

\begin{abstract}
  We briefly explain the structure of the Hartree-Fock-Bardeen-Cooper-Schrieffer (HFBCS) and Quasi-particle Random Phase Approximaiton (QRPA) code {\tt hfbcs-qrpa}, highlighting only the differences with the code {\tt skyrme\_rpa} that has been previously published \cite{CCVGC:2013}. The code deals with open shell spherical systems and non-spin flip, non-charge-exchange and natural parity $(-1)^l$ excitations.  
\end{abstract}


\maketitle

\section{Differences with the published code {\tt skyrme\_rpa}}

The present code is an extension of the Hartree-Fock plus Random Phase Approximation (HF plus RPA) code that has 
already been published in Ref. \cite{CCVGC:2013}. HF is extended to HF-Bardeen-Cooper-Schrieffer (HF-BCS) and,
accordingly, RPA to Quasiparticle RPA (QRPA). The scheme remains self-consistent, once the effective pairing force
is introduced.

Along the code we use $\hbar c=197.327053$ MeV fm and $m c^2= 938.91869$ MeV, obtained as $(m_p + m_n)/2$, for both 
neutrons and protons.      

\subsection{HF-BCS}

HF equations are solved in coordinate space, assuming spherical symmetry, as explained in \cite{CCVGC:2013}, 
and employing the standard form of a Skyrme interaction\footnote{In the current version, tensor terms \cite{stancu1977} 
are implemented at the HF-BCS level but not in the residual QRPA interaction.}. Available Skyrme interactions are: SkM*, SkP, SLy4, SLy5, SkX, KDE0-J33 (KDE33) and SAMi. The HF equations are coupled to the standard BCS equations, that in spherical symmetry with $\alpha=(n,l,j)$ read
\begin{eqnarray}
&& \sum_\alpha (2j_\alpha+1) v_\alpha^2 = N, \hfill \label{number} \\
&& \Delta_\alpha = -\sum_{\alpha'} \frac{\Delta_{\alpha'}}{2E_{\alpha'}}V_{\alpha\tilde\alpha \alpha' 
\tilde \alpha'}, \label{gap}
\end{eqnarray}
where $E$, $u$ and $v$ are the usual quasi-particle energies and BCS amplitudes as defined in the 
literature \cite{RS.80}, and the tilde denotes the time-reversal state. 
$N$ is the particle number. Either proton pairing, or neutron pairing, or both, can be active. 
No neutron-proton pairing is included and, thus, the particles retain their quantum number $t_z$.

To keep $u$ and $v$ positive, we use the phase convention in which the single-particle (s.p.) wave functions include a factor $i^l$ [Eq. (7) of Ref. 
\cite{CCVGC:2013} should be modified accordingly].
The number and gap equations, (\ref{number}) and (\ref{gap}), are solved in a limited window which is specified by the user if the pairing is run in manual mode. In that case, an upper limit on the corresponding s.p. energy $\varepsilon_\alpha$ and a fixed number of maximum s.p. levels above the Fermi energy to be considered should be specified. Note that only one of those two pairing cutoffs, namely the lowest, will act as the real one. These two possibilities give more flexibility in the usage of the code. Both types of cutoffs can be different for neutrons and protons. There is no lower limit imposed for this pairing window. For those users not conversant with pairing calculations or not interested in exploring pairing effects, we strongly recommend to use pairing in automatic mode, in which case, a volume pairing force (see below) has been fitted to reproduce the experimental neutron pairing gap in ${}^{120}$Sn ($\Delta=1.245$ MeV) with a neutron pairing window fixed by taking into account only bound states above the neutron Fermi energy and to a maximum of 6 neutron s.p. levels. The default pairing window in automatic mode is set in an analogous way.    

The matrix elements $V_{\alpha\tilde\alpha \alpha' \tilde \alpha'}$ are calculated using a zero-range, density-dependent
pairing force of the type
\begin{eqnarray}
V(\textbf{r}_{1},\textbf{r}_{2}) & = & V_{0}\left[1 + x
\left(
\frac{\rho(\frac{\textbf{r}_{1}+\textbf{r}_{2}}{2})}{\rho_{0}}
\right)
\right] \delta(\textbf{r}_{1}-\textbf{r}_{2}) 
\hfill \nonumber \\
& \equiv & V_{\rm pair}(r)\delta(\textbf{r}_{1}-\textbf{r}_{2}). 
\label{vpair}
\end{eqnarray}
In the above equation, $\rho_{0}=0.16$ fm$^{-3}$ while the other two parameters $x$ and $V_0$ have to be adjusted. $x = 0, 1$ and 
$0.5$ define the usual volume, surface and mixed pairing. $V_0$ is usually adjusted to reproduce some given pairing gap but it has to be noted that the code does not make this adjustement by itself unless automatic mode is choseen where only the use of volume pairing ($x=0$) is allowed. 

The total HF-BCS energy can be calculated, and is calculated by the code, in two different manners.
It can be calculated from the force, or energy functional, directly. In this case
\begin{equation}\label{eq:E1}
E = E_{\rm kin} + E_{\rm Skyrme} + E_{\rm Coul} + E_{\rm pair}.
\end{equation}
The first two terms are the kinetic and Skyrme contributions to the energy and are written as the
volume integrals of the corresponding energy densities, ${\cal E}_{\rm kin}$ and ${\cal E}_{\rm Skyrme}$, 
as in Eqs. (2.1) and (2.2) of Ref. \cite{chabanat1998}, or in Eqs. (24) and (38) of Ref. \cite{ryssens}. 
The Coulomb energy has a direct part and an exchange part that is calculated in the usual Slater
approximation. The pairing energy is written as in Eq. (16) of \cite{ryssens}.
As an alternative,
\begin{equation}\label{eq:E2}
E = \frac{1}{2}\sum_\alpha v_\alpha^2\varepsilon_\alpha+
\frac{1}{2}E_{\rm kin} + E_{\rm rearr} + E_{\rm pair},
\end{equation}
where the first term includes the s.p. energies $\varepsilon_\alpha$, and $E_{\rm rearr}$ is the rearrangement
term associated with the density-dependent term in the Skyrme force (as
in Eq. (109) of \cite{ryssens}), as well as with the Coulomb
exchange energy within the Slater approximation.
The equality between the two expressions (\ref{eq:E1}) and (\ref{eq:E2}) for the total energy is,
among the rest, a useful test for the convergence of the HF-BCS equations.


Regarding the calculation of charge radii, we take into account the electromagnetic finite size of the proton and the neutron. That is, we assume the proton and neutron electric sizes as $\langle r_p^2\rangle = 0.64$ fm$^2$ and $\langle r_n^2\rangle = -0.11$ fm$^2$, respectively, as well as the anomalous proton magnetic moment $\kappa_p=\mu_p-1$ with $\mu_p=2.79$ and neutron magnetic moment $\kappa_n=\mu_n=-1.91$. The expression for the charge radius $\langle r^2\rangle_{\rm ch}$ used is derived in the non-relativistic limit and can be written as \cite{horowitz2012} (see also \cite{chabanat1997}),
\begin{eqnarray}\label{eq:rch}
  \langle r^2\rangle_{\rm ch}&=&\langle r^2\rangle_{\rm p}+\langle r_p^2\rangle+\frac{N}{Z}\langle r_n^2\rangle \nonumber \\
  &&+\frac{1}{Z}\left(\frac{\hbar}{m c}\right)^2\sum_{\alpha\tau}v^2_{\alpha\tau}(2j_\alpha+1)\kappa_\tau\langle{\bm \sigma}\cdot {\bm l}\rangle_\alpha \ ,
\end{eqnarray}
where $\tau=p,~n$ indicates if the nucleon is a proton or a neutron, respectively, and $\langle r^2\rangle_{\rm p}$ is the mean square radius of the density distribution of protons,
\begin{equation}\label{eq:rp}
\langle r^2\rangle_{\rm p}=\int d{\bm r} r^2\rho_p({\bm r}) \ .
\end{equation}

We also evaluate the isospin mixing $\varepsilon$ in the obtained s.p. wave-functions as discussed in Sec. IIA of Ref.~\cite{roca-maza2020}. Specifically,   
\begin{equation}
\varepsilon^2 \equiv \frac{1}{N-Z+2}\sum_{\substack{n_p,n_n\\l_p=l_n\\j_p=j_n}}(2j_p+1)v_p^2u_n^2\mathcal{O}_{np}^2
\label{epsilon}
\end{equation}
where we have omitted the $\alpha$ label (previously defined) for the sake of simplicity and clarity. The overlap factor $\mathcal{O}_{np}$ between the neutron ($n$) and proton ($p$) radial part $R_{n,l}(r)$ of the considered s.p. wave function is defined as 
\begin{equation}
  \mathcal{O}_{np} \equiv \int_0^\infty dr r^2 R_{n_p,l_p}(r) R_{n_n,l_n}(r) \ .
\label{gamma}
\end{equation}

\subsection{QRPA}

The QRPA equations are
\begin{eqnarray}
\left(
\begin{array}{c c}
A_{\alpha\beta,\gamma\delta}         & B_{\alpha\beta,\gamma\delta}         \\
-B^{\ast}_{\alpha\beta,\gamma\delta} & -A^{\ast}_{\alpha\beta,\gamma\delta} \\
\end{array}
\right)
\left(
\begin{array}{c}
X^{\nu}_{\gamma\delta}  \\ Y^{\nu}_{\gamma\delta} \\
\end{array}
\right)
&=&
E_{\nu}
\left(
\begin{array}{c}
X^{\nu}_{\alpha\beta}  \\ Y^{\nu}_{\alpha\beta} \\
\end{array}
\right).
\end{eqnarray}
The basis on which these equations are written is made up with two quasiparticle states, like 
$(\alpha\beta)$ and $(\gamma\delta)$. These respect the usual selection rules in that they
are coupled to good angular momentum $J$ and parity $\pi$. The upper limit on the two
quasiparticle energy can be specified (see below). On the other hand, the code
automatically discards configurations for which both quasiparticle states are fully occupied or both have an occupation probability lower than $10^{-6}$. 

\begin{widetext}
The matrix elements $A$ and $B$, in the angular momentum-coupled representation, written on the HF-BCS basis
two-quasiparticle basis, have the form
\begin{eqnarray}
A_{\alpha\beta,\gamma\delta}
&=&
\frac{1}{\sqrt{1+\delta_{\alpha\beta}}\sqrt{1+\delta_{\gamma\delta}}}
\nonumber\\
&&\times
\left[
\left( E_\alpha + E_\beta \right)\delta_{\alpha\gamma}\delta_{\beta\delta} 
+ G_{\alpha\beta\gamma\delta}
(u_{\alpha}u_{\beta}u_{\gamma}u_{\delta} + v_{\alpha}v_{\beta}v_{\gamma}v_{\delta})
\right.
\nonumber\\
&&
\left.
+ F_{\alpha\beta\gamma\delta}
(u_{\alpha}v_{\beta}u_{\gamma}v_{\delta} + v_{\alpha}u_{\beta}v_{\gamma}u_{\delta})
- (-1)^{j_{\gamma}+j_{\delta}-J^{\prime}}F_{\alpha\beta\delta\gamma}
(u_{\alpha}v_{\beta}v_{\gamma}u_{\delta} + v_{\alpha}u_{\beta}u_{\gamma}v_{\delta})
\right], 
\\
B_{\alpha\beta,\gamma\delta}
&=&
\frac{1}{\sqrt{1+\delta_{\alpha\beta}}\sqrt{1+\delta_{\gamma\delta}}}
\nonumber\\
&&\times
\left[
- G_{\alpha\beta\delta\gamma}
(u_{\alpha}u_{\beta}v_{\gamma}v_{\delta} + v_{\alpha}v_{\beta}u_{\gamma}u_{\delta})
- (-1)^{j_{\delta}+j_{\gamma}-J^{\prime}}F_{\alpha\beta\delta\gamma}
(u_{\alpha}v_{\beta}u_{\gamma}v_{\delta} + v_{\alpha}u_{\beta}v_{\gamma}u_{\delta})
\right.
\nonumber\\
&&
\left.
+ (-1)^{j_{\alpha}+j_{\beta}+j_{\gamma}+j_{\delta}-J-J^{\prime}}F_{\alpha\beta\gamma\delta}
(u_{\alpha}v_{\beta}v_{\gamma}u_{\delta} + v_{\alpha}u_{\beta}u_{\gamma}v_{\delta})
\right],
\end{eqnarray}
with
\begin{eqnarray}
G_{\alpha\beta\gamma\delta}
&=&
\sum_{m_{\alpha}m_{\beta}m_{\gamma}m_{\delta}}
\langle j_{\alpha}m_{\alpha}j_{\beta}m_{\beta}|JM\rangle
\langle j_{\gamma}m_{\gamma}j_{\delta}m_{\delta}|J^{\prime}M^{\prime}\rangle
V^{pp}_{\alpha\beta,\gamma\delta}, 
\\
F_{\alpha\beta\gamma\delta}
&=&
\sum_{m_{\alpha}m_{\beta}m_{\gamma}m_{\delta}}
\langle j_{\alpha}m_{\alpha}j_{\beta}m_{\beta}|JM\rangle
\langle j_{\gamma}m_{\gamma}j_{\delta}m_{\delta}|J^{\prime}M^{\prime}\rangle
V^{ph}_{\alpha\bar{\delta}\bar{\beta}\gamma}. \label{meph}
\end{eqnarray}
\end{widetext}
$V^{ph}_{\alpha\bar{\delta}\bar{\beta}\gamma}$ are the (uncoupled) matrix elements
of the particle-hole effective interaction, and $V^{pp}_{\alpha\beta,\gamma\delta}$
represent the (uncoupled) matrix elements of the particle-particle effective interaction.
The p-h effective interaction is described in detail in Ref. \cite{CCVGC:2013}. 
In fact, one can immediately spot that the matrix element in Eq. (\ref{meph}) 
is the same that has been defined in Eq. (16) of Ref. \cite{CCVGC:2013}; the different notation
here has the purpose of distinguishing it clearly from the matrix element $G$ that is the one
of the pairing force and reads 
\begin{widetext}
\begin{equation}
G_{\alpha\beta\gamma\delta} = R \sum_\lambda \left( (-)^{j_\beta+j_\gamma+J} 
\left\{ \begin{array}{ccc} j_\alpha & j_\gamma & \lambda \\ j_\delta & j_\beta & J \end{array} \right\} 
\langle \alpha \vert\vert Y_\lambda \vert\vert \gamma \rangle
\langle \beta \vert\vert Y_\lambda \vert\vert \delta \rangle +
(-)^{j_\beta+j_\gamma} 
\left\{ \begin{array}{ccc} j_\alpha & j_\delta & \lambda \\ j_\gamma & j_\beta & J \end{array} \right\} 
\langle \alpha \vert\vert Y_\lambda \vert\vert \delta \rangle
\langle \beta \vert\vert Y_\lambda \vert\vert \gamma \rangle \right),
\end{equation}
\end{widetext}
where $R$ is the radial integral
\begin{equation}
R = \int \frac{dr}{r^2} u_\alpha(r)u_\beta(r)u_\gamma(r)u_\delta(r)V_{\rm pair}(r).
\end{equation}

The excitation operators and associated sum rules used in the code are those discussed in Ref.~\cite{CCVGC:2013}. Regarding spurious states: the HF-BCS is known to break different symmetries present in the original Hamiltonian. In principle QRPA should restore those symmetries exactly. However, numerical implementations are not exact, and spurious states do not appear necessarily at zero energy as expected. As a rule, the sppurious state should be the first excited state, close to zero energy, in the calculated spectrum. Specifically, within the present code: i) for 0$^+$ excitations, we have not implemented any projection technique to avoid the spurious state originated from the violation of the particle number; and ii) for 1$^-$ excitations, we use the operators defined in \cite{CCVGC:2013} to approximately remove the spurous state that arise from the violation of translational invariance.

\onecolumngrid

\section{Input of the code}
In this section we briefly describe the input file {\tt hfbcs-qrpa.in} needed to run the code. The format of the input file should be as follows: 
\begin{verbatim}
Line Input         Explanation
---- -----         -----------
 1   FORCE           Name of the Skyrme interaction to be used 
                     [options: SkM*, SkP, SLy4, SLy5, SkX, KDE33, SAMi]
 2   MESH STEP       Number of mesh points and step (fm) 
 3   A Z             Mass number (A) and proton number (Z)
 4   EPS             Required accuracy for the HFBCS equations. Specifically, it corresponds 
                     to the maximum accepted difference between consecutive iterations and it 
                     is used for: single-particle energy, gap and chemical potential.                
 5   FLAG BCSP BCSN  Automatic (0) or manual(1) BCS calculaitons. Active (1) non-active (0) 
                     for protons. Active (1) non-active (0) for neutrons. 
                     Only if manual mode has been chosen next three lines (6-8) should be added:    
 6   LEVP LEVN       Number of levels taken above the Fermi level in the BCS calculation for 
                     protons and neutrons. 
 7   EMAXP EMAXN     Maximum proton and neutron energies for the BCS calculation. 
 8   -V_0 x          Parameters of the pairing interaction to be used defined in 
                     the previous section.         
 9   CALEX           Type of calculations [options: 
                     - UNP: stands for unperturbed calculation, residual interaction set to zero  
                     - TDA: Tamm-Dancoff Approximation is used
                     - RPA: Random Pahse Approximation is used]. 
10   J  P            Total angular momentum and pairty (+1 or -1). At the moment only natural 
                     parity is considered. 
11   ECP ECN         Minimum and maximum energy for the considered two body 
                     neutron-neutron and proton-proton basis states.
\end{verbatim}

\section{Output of the code}
The code produces different output files. Definitions of the different quantities such as operators or transition densities are as in {\sc skyrme\_rpa} \cite{CCVGC:2013}. Below a brief description:  

\begin{verbatim}
File name                   Description
---------                   -----------
hfbcs-qrpa.out              main output of the code                     
quasiparticle_states.out    full list of quasi-particle states used in the calculation.
                            Information on the Fermi energies is also given     
basis_states.out            full list of two-body neutron-neutron and proton-proton 
                            basis states used in the calculation.       
d.out                       proton and neutron densities 
td.out                      proton and neutron transition densities for all excited states
strength_*.out              three files contain isoscalar, isovector and electromagnetic 
                            strength functions (reduced transition probabilities convoluted 
                            by a Lorenzian function with a energy step of 0.1 MeV and width 
                            of 1 MeV)
\end{verbatim}
In the main output file ({\tt hfbcs-qrpa.out}) details on the parameters of the used functional as well as active terms are printed. One-body center of mass correction is active for all the Skyrme interaction assuming the same prescription as in {\sc skyrme\_rpa}. Convergence information and a list of quasi-particle levels is given. Total binding energy as well as its different contributions are printed in two different forms (see the discussion above). Firstly, the energy is calculated as the integral of the energy density. Then, the energy is calculated via the sum of the eigenvalues. In this latter case the rearrangement energy is needed. The calculation of the binding energy in two different ways serves as a strong test for the convergence the code.
Neutron and proton root mean square radii are also printed and the root mean square radius is estimated. Average pairing gaps, and isospin mixing in the groud state wave function are also given in the output. Finally, different sum rules and isoscalar, isovector and electromagnetic reduced transition probabilities are listed.   

\section{Example} 
In this section, we show an input ({\tt hfbcs-qrpa.in}) example of the code which uses the manual pairing option (FLAG=1):    
\begin{verbatim}
SLy5                   
200 0.1
120 50
1.d-6
1 0 1
0 6
0. 0.
870.6 1.
RPA
2 +1
0. 100.
\end{verbatim}
The main output ({\tt hfbcs-qrpa.out}) of the code corresponding to this input is given below. 
\begin{verbatim}
  . . .
  
 >>>>>>>>>>>>>>>>>>>>>>>>>>>>>>>>>>>>>>>>>>>>>>>
      READING SKYRME INTERACTION TO BE USED     
 >>>>>>>>>>>>>>>>>>>>>>>>>>>>>>>>>>>>>>>>>>>>>>>
 
 NAME: SLy5 
 
 SKYRME FORCE PARAMETERS IN STANDARD FORM:
 
           t0  =   -2484.88000
           x0  =       0.77800
           t1  =     483.13000
           x1  =      -0.32800
           t2  =    -549.40000
           x2  =      -1.00000
           t3  =   13763.00000
           x3  =       1.26700
           g   =       0.16667
           W0  =     126.00000
           W0p =     126.00000
 
 PARAMETERS OF THE SKYRME ENERGY DENSITY FUNCTIONAL:
 as defined in J. Dobaczewski and J. Dudek,
 Phys. Rev. C52, 1827 (1995)
 
      C_0^\rho =    -931.83000 +     860.18750 *\rho^ 0.16667
      C_1^\rho =     793.91916 +   -1013.30087 *\rho^ 0.16667
C_0^\delta\rho =     -76.52453
C_1^\delta\rho =      16.37485
      C_0^\tau =      56.24937
      C_1^\tau =      23.95020
         C_0^s =    -172.69916 +     439.84254 *\rho^ 0.16667
         C_1^s =     310.61000 +    -286.72917 *\rho^ 0.16667
  C_0^\nabla J =     -94.50000
  C_1^\nabla J =     -31.50000
  C_0^\Delta s =      46.08734
  C_1^\Delta s =      14.06234
         C_0^T =     -15.66645
         C_1^T =     -64.53312
 
 >>>>>>>>>>>>>>>>>>>>>>>>>>>>>>>>>>>>>>>>>>>>>>>
          HF+BCS CALCULATION IN A BOX           
 >>>>>>>>>>>>>>>>>>>>>>>>>>>>>>>>>>>>>>>>>>>>>>>
 
 NUCLEUS:
 -Mass number   120.
 -Proton number  50.
 
 BOX PARAMETERS:
 -N. of points   200
 -Step size       0.10E+00
 
 CONVERGENCE PARAMETERS:
 -Error spe       0.10E-05
 -Error gap eq.   0.10E-05
 -Error num. eq.  0.10E-05
 
 POTENTIAL INCLUDED IN THE CALCULATIONS:
 -SKYRME STANDARD WITH
       |___ J2 TERMS... YES
       |___ TENSOR..... NO
 -COULOMB DIRECT....... YES
 -COULOMB EXCHANGE..... YES
 -SPIN-ORBIT........... YES
 -DEFAULT PAIRING...... NO
 -PROT. PAIRING (BCS).. NO
 -NEUT. PAIRING (BCS).. YES
        ENERGY CUT.....  0.000000000000000E+000
        LEV. ABOVE F...           6
    Type: delta force      
    Vp= V0(1-x(rho/rho0))^g
   V0  =   0.87060E+03
   x   =   0.10000E+01
   rho0=   0.16000E+00
   g   =   0.10000E+01
 -CENTER OF MASS CORR.. YES
 
 CONVERGENCE:
 -At iteration number:             77
 -Maximum  error in spe:  0.99830E-05
 
 >>>>>>>>>>>>>>>>>>>>>>>>>>>>>>>>>>>>>>>>>>>>>>>
 SINGLE PARTICLE ENERGIES, OCCUP., V^2 AND GAPS 
 >>>>>>>>>>>>>>>>>>>>>>>>>>>>>>>>>>>>>>>>>>>>>>>
 
 PROTON STATES
 
   1S1/2     E=-0.4664E+02     DEG= 0.2000E+01     V^2= 0.1000E+01      D=  0.0000
   1P3/2     E=-0.3873E+02     DEG= 0.4000E+01     V^2= 0.1000E+01      D=  0.0000
   1P1/2     E=-0.3778E+02     DEG= 0.2000E+01     V^2= 0.1000E+01      D=  0.0000
   1D5/2     E=-0.2980E+02     DEG= 0.6000E+01     V^2= 0.1000E+01      D=  0.0000
   1D3/2     E=-0.2756E+02     DEG= 0.4000E+01     V^2= 0.1000E+01      D=  0.0000
   2S1/2     E=-0.2559E+02     DEG= 0.2000E+01     V^2= 0.1000E+01      D=  0.0000
   1F7/2     E=-0.2040E+02     DEG= 0.8000E+01     V^2= 0.1000E+01      D=  0.0000
   1F5/2     E=-0.1634E+02     DEG= 0.6000E+01     V^2= 0.1000E+01      D=  0.0000
   2P3/2     E=-0.1522E+02     DEG= 0.4000E+01     V^2= 0.1000E+01      D=  0.0000
   2P1/2     E=-0.1361E+02     DEG= 0.2000E+01     V^2= 0.1000E+01      D=  0.0000
   1G9/2     E=-0.1083E+02     DEG= 0.1000E+02     V^2= 0.1000E+01      D=  0.0000
 
 NEUTRON STATES
 
   1S1/2     E=-0.5616E+02     DEG= 0.2000E+01     V^2= 0.1000E+01      D=  0.2017
   1P3/2     E=-0.4748E+02     DEG= 0.4000E+01     V^2= 0.1000E+01      D=  0.4678
   1P1/2     E=-0.4615E+02     DEG= 0.2000E+01     V^2= 0.1000E+01      D=  0.4191
   1D5/2     E=-0.3787E+02     DEG= 0.5999E+01     V^2= 0.9998E+00      D=  0.7963
   1D3/2     E=-0.3481E+02     DEG= 0.3999E+01     V^2= 0.9998E+00      D=  0.7045
   2S1/2     E=-0.3334E+02     DEG= 0.1999E+01     V^2= 0.9997E+00      D=  0.8519
   1F7/2     E=-0.2775E+02     DEG= 0.7993E+01     V^2= 0.9991E+00      D=  1.1555
   1F5/2     E=-0.2250E+02     DEG= 0.5992E+01     V^2= 0.9987E+00      D=  1.0469
   2P3/2     E=-0.2194E+02     DEG= 0.3993E+01     V^2= 0.9982E+00      D=  1.1819
   2P1/2     E=-0.2013E+02     DEG= 0.1995E+01     V^2= 0.9975E+00      D=  1.1866
   1G9/2     E=-0.1747E+02     DEG= 0.9937E+01     V^2= 0.9937E+00      D=  1.4813
   2D5/2     E=-0.1137E+02     DEG= 0.5787E+01     V^2= 0.9644E+00      D=  1.2543
   1G7/2     E=-0.9976E+01     DEG= 0.7133E+01     V^2= 0.8916E+00      D=  1.3924
   3S1/2     E=-0.9077E+01     DEG= 0.1618E+01     V^2= 0.8089E+00      D=  1.0877
   2D3/2     E=-0.8573E+01     DEG= 0.2530E+01     V^2= 0.6325E+00      D=  1.2771
  1H11/2     E=-0.7227E+01     DEG= 0.2978E+01     V^2= 0.2481E+00      D=  1.7057
   2F7/2     E=-0.1526E+01     DEG= 0.4533E-01     V^2= 0.5666E-02      D=  1.0167
   3P3/2     E=-0.1663E-01     DEG= 0.2758E-02     V^2= 0.6894E-03      D=  0.4313
   2F5/2     E= 0.1169E+01     DEG= 0.0000E+00     V^2= 0.0000E+00      D=  0.0000
   1H9/2     E= 0.1856E+01     DEG= 0.0000E+00     V^2= 0.0000E+00      D=  0.0000
  1I13/2     E= 0.2809E+01     DEG= 0.0000E+00     V^2= 0.0000E+00      D=  0.0000
 
 >>>>>>>>>>>>>>>>>>>>>>>>>>>>>>>>>>>>>>>>>>>>>>>
    CONTRIBUTIONS TO THE TOTAL ENERGY (MeV):
 >>>>>>>>>>>>>>>>>>>>>>>>>>>>>>>>>>>>>>>>>>>>>>>
 
 -INTEGRAL OF THE ENERGY DENSITY:
  E(KIN) =          0.21769E+04
  E(SKYRME) =      -0.34793E+04
   |___E(t0) =     -0.12698E+05
   |___E(t3) =      0.82849E+04
   |___E(t1,t2) =   0.93408E+03
  E(SO) =          -0.54885E+02
  E(CD) =           0.36706E+03
  E(CE) =          -0.19131E+02
  E(pair) =        -0.99067E+01
  E(TOT) =         -0.10193E+04
 
 -HF ENERGY + E_REA:
  E(KIN) =          0.10885E+04
  E(SPE) =         -0.14010E+04
  E(REA) =         -0.69678E+03
  E(pair) =        -0.99067E+01
  E(TOT) =         -0.10193E+04
 
 -NUMERICAL CHECKS:
  E_INT/E_HF-1    
  REL. ERR. (%) =   0.36779E-03
 
 >>>>>>>>>>>>>>>>>>>>>>>>>>>>>>>>>>>>>>>>>>>>>>>
             ROOT MEAN SQUARE RADII
 >>>>>>>>>>>>>>>>>>>>>>>>>>>>>>>>>>>>>>>>>>>>>>>
 
  R(N) =            0.47278E+01
  R(P) =            0.45865E+01
  R(CH) =           0.46418E+01
  R^2(CH) =         0.21546E+02
   |___R^2(p) =     0.64000E+00
   |___N*R^2(n)/Z= -0.15400E+00
   |___R^2(SO-n) = -0.38991E-01
   |___R^2(SO-p) =  0.63250E-01
 
 >>>>>>>>>>>>>>>>>>>>>>>>>>>>>>>>>>>>>>>>>>>>>>>
          AVERAGE PAIRING GAPS (MeV):
 >>>>>>>>>>>>>>>>>>>>>>>>>>>>>>>>>>>>>>>>>>>>>>>
 
  GAP(P) =          0.00000E+00
  GAP(n) =          0.14505E+01
 
 
 >>>>>>>>>>>>>>>>>>>>>>>>>>>>>>>>>>>>>>>>>>>>>>>
          ISOSPIN MIXING IN THE HF G.S.         
    |HF> = sqrt(1-e^2)|T0, T0> + e|T0+1, T0>    
 >>>>>>>>>>>>>>>>>>>>>>>>>>>>>>>>>>>>>>>>>>>>>>>
 
  e^2 (%) =         0.31506E+00
 
 
 >>>>>>>>>>>>>>>>>>>>>>>>>>>>>>>>>>>>>>>>>>>>>>>
         LINEAR RESPONSE EVALUATED IN           
          RANDOM PHASE APPROXIMATION            
 >>>>>>>>>>>>>>>>>>>>>>>>>>>>>>>>>>>>>>>>>>>>>>>
 
 
 CONSIDERED TRANSITIONS WITH:
   J                          2
   PARITY                     1
 
 >>>>>>>>>>>>>>>>>>>>>>>>>>>>>>>>>>>>>>>>>>>>>>>
        MOMENTS OF THE STRENGTH FUNCTION        
 >>>>>>>>>>>>>>>>>>>>>>>>>>>>>>>>>>>>>>>>>>>>>>>
 
 -NON-ENERGY WEIGHTED SUM RULE:
  m(0)(IS)  =       0.26417E+05
  m(0)(IV)  =       0.11253E+05
 
 -INVERSE ENERGY WEIGHTED SUM RULE:
  m(-1)(IS) =       0.10068E+05
  m(-1)(IV) =       0.14632E+04
 
 -ENERGY WEIGHTED SUM RULE:
  m( 1)(IS) =       0.20875E+06
  m( 1)(IV) =       0.23841E+06
  m(1)[D.C.]IS =    0.21407E+06
  Exhaus. IS(%)=       97.51375
 IV DOUBLE COMMUTATOR EWSR NOT AVAILABLE
 
 -ENERGY CUBE WEIGHTED SUM RULE:
  m( 3)(IS) =       0.42103E+08
  m( 3)(IV) =       0.16041E+09
 
 >>>>>>>>>>>>>>>>>>>>>>>>>>>>>>>>>>>>>>>>>>>>>>>
          AVERAGE EXCITATION ENERGIES           
 >>>>>>>>>>>>>>>>>>>>>>>>>>>>>>>>>>>>>>>>>>>>>>>
  E_CONSTR. (IS) =  0.45533E+01
  E_CENTR. (IS) =   0.79021E+01
  E_SCAL. (IS) =    0.14202E+02
  E_CONSTR. (IV) =  0.12765E+02
  E_CENTR. (IV) =   0.21187E+02
  E_SCAL. (IV) =    0.25939E+02
 
 >>>>>>>>>>>>>>>>>>>>>>>>>>>>>>>>>>>>>>>>>>>>>>>
        REDUCED TRANSITION PROBABILITIES        
 >>>>>>>>>>>>>>>>>>>>>>>>>>>>>>>>>>>>>>>>>>>>>>>
 
      E            BEL_is       FRAC_NEWSR   BEL_em       BEL_iv       FRAC_NEWSR
    ------------------------------------------------------------------------------
    1 0.13051E+01  0.11155E+05  0.42229E+02  0.12264E+04  0.12658E+04  0.11249E+02
    2 0.27263E+01  0.29313E+02  0.11096E+00  0.41912E+00  0.16969E+02  0.15080E+00
    3 0.30799E+01  0.57219E+03  0.21660E+01  0.87101E+02  0.27614E+02  0.24539E+00
    4 0.36154E+01  0.59934E+03  0.22688E+01  0.63858E+02  0.72237E+02  0.64195E+00
    5 0.43815E+01  0.33156E+02  0.12551E+00  0.96100E+01  0.19523E+00  0.17350E-02
    6 0.46192E+01  0.19115E+02  0.72361E-01  0.62297E+01  0.38413E+00  0.34136E-02
    7 0.48144E+01  0.68087E+01  0.25774E-01  0.13742E+01  0.70107E-01  0.62302E-03
    8 0.56035E+01  0.10373E+04  0.39268E+01  0.48223E+03  0.13716E+03  0.12189E+01
    9 0.56320E+01  0.37730E+03  0.14283E+01  0.18057E+03  0.55514E+02  0.49333E+00
   10 0.66789E+01  0.11180E+02  0.42321E-01  0.14908E+00  0.66120E+01  0.58758E-01
    
                                          . . .
   
 1080 0.98378E+02  0.28779E-05  0.10894E-07  0.14364E-02  0.60056E-02  0.53369E-04
 1081 0.98572E+02  0.27497E-04  0.10409E-06  0.21067E-02  0.74915E-02  0.66575E-04
 1082 0.98620E+02  0.10907E-04  0.41290E-07  0.41446E-04  0.26174E-03  0.23260E-05
 1083 0.99015E+02  0.52814E-04  0.19993E-06  0.38653E-04  0.26697E-04  0.23725E-06
 1084 0.99206E+02  0.33656E-03  0.12741E-05  0.80204E-02  0.25846E-01  0.22969E-03
 1085 0.99424E+02  0.72734E-03  0.27533E-05  0.35527E-04  0.22645E-03  0.20124E-05
 1086 0.99529E+02  0.44898E-01  0.16996E-03  0.14265E-04  0.48156E-01  0.42795E-03
 1087 0.99840E+02  0.12720E-04  0.48151E-07  0.12853E-02  0.46425E-02  0.41256E-04
 1088 0.99994E+02  0.68463E-04  0.25917E-06  0.19422E-02  0.92958E-02  0.82608E-04
 1089 0.10014E+03  0.25707E-02  0.97315E-05  0.10805E-02  0.22619E-03  0.20101E-05
 1090 0.10151E+03  0.25770E-02  0.97551E-05  0.44132E-04  0.14046E-02  0.12482E-04
 
   CPU TIME IN SECONDS =  2017.41
\end{verbatim}

\begin{figure}[h!]
\vspace{5mm}  
\includegraphics[width=0.6\linewidth,clip=true]{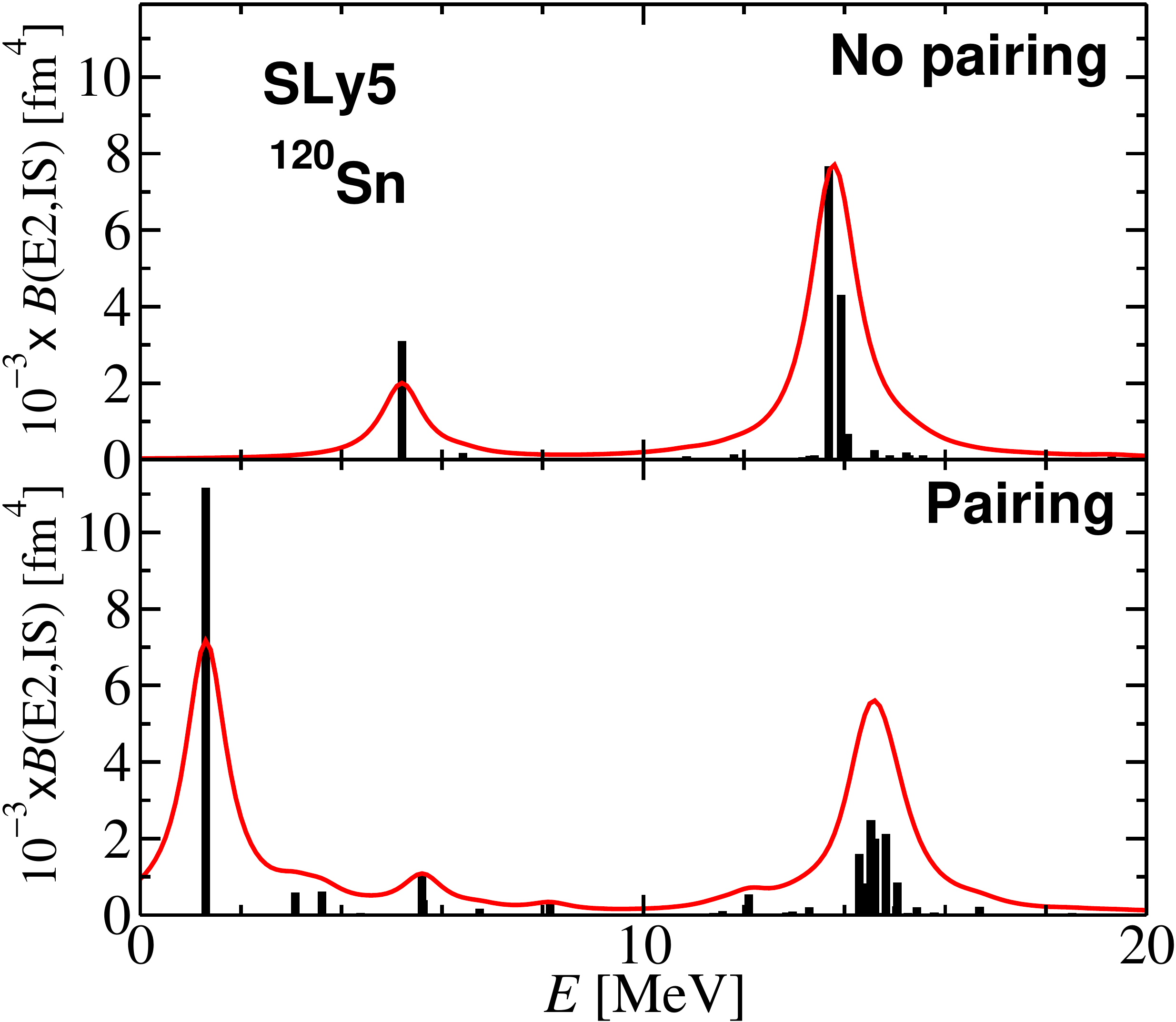}
\caption{Reduced transition probabilities (bars) for the text example contained in {\tt hfbcs-qrpa.out} and their convolution with a Lorenzian function (solid lines) with a width of 1 MeV. The latter information is also given by the code and can be found in the file {\tt strength\_is.out} for the isoscalar response (upper panel) and {\tt strength\_iv.out} for the isovector response (lower panel).}
\label{fig1}
\end{figure}

Finally, for this example, we show in Fig.~\ref{fig1} the reduced transition probabilities contained in {\tt hfbcs-qrpa.out} and their convolution with a Lorenzian function with a width of 1 MeV. The latter information is also given by the code and can be found in the file {\tt strength\_is.out} for the isoscalar response and {\tt strength\_iv.out} for the isovector response.

\bibliography{bibliography}

\end{document}